\documentclass{PoS}

\title{Dynamical simulations with HYP-link Wilson fermions}

\ShortTitle{Dynamical simulations with HYP-link Wilson fermions}

\author{\speaker{Stefan Schaefer}\\%
        NIC, DESY,
	Platanenallee 6,
	D-15738 Zeuthen,
	Germany\\
	E-mail: \email{stefan.schaefer@desy.de}}

\author{Anna Hasenfratz,  Roland Hoffmann\\
        Department of Physics, University of Colorado,
	Boulder, CO 80309 USA\\
        E-mail: \email{anna@pizero.colorado.edu},  \email{hoffmann@pizero.colorado.edu}}

\abstract{We present results from simulations of two dynamical flavors of
          improved Wilson fermions with nHYP smeared gauge links. We demonstrate
	  that the simulation is stable at a pseudo-scalar mass of 360MeV, a 
	  2.1fm box and a lattice spacing of 0.13fm.}

\FullConference{The XXV International Symposium on Lattice Field Theory\\
		July 30 - August 4 2007\\
   	         Regensburg, Germany}

\begin{document}

\section{Introduction}
Ever since the introduction of the original APE smearing in 1987\cite{Albanese:1987ds},
many applications have shown the beneficial effects of constructing the lattice Dirac 
operator from smeared  links.  The hypothesis behind employing smeared links is that 
short range fluctuations of the gauge field, the so-called dislocations, are responsible 
for some of the poor scaling behavior and large cost of fermion simulations. For 
different lattice discretizations of the Dirac operator, the improvement due to smearing 
comes in different disguises.  In Wilson fermions dislocations cause the exceptionally 
small eigenvalues of the Dirac operator, in staggered
fermions the large taste breaking. Domain Wall fermions acquire their residual
explicit chiral symmetry breaking through dislocations and they are responsible
for the high numerical cost of constructing the overlap operator. 

An alternative approach to reduce these effects is the use of special
gauge actions, e.g. the Iwasaki action or DBW2, which suppress the occurrence
of dislocations. However, it has turned out that these gauge actions themselves
induce quite large scaling violations. Smearing the links, however, reduces the 
effect of the  dislocations on the fermions only. They remain part of the gauge 
dynamics and one can choose a gauge action which does not introduce poor scaling
behavior from the gluonic sector.
Increased auto-correlation times in molecular dynamics based 
algorithms have been observed when using improved gauge actions
as well. Below, we will show that we do not observe signs of this in our simulations.

Regardless of whether one accepts the explanation of these effects, in many quenched 
studies it was demonstrated that smearing helps to improve scaling in the situations 
listed above. It is therefore natural to use it in dynamical simulations too.

The goal of this conference contribution is to convince the reader that simulations
of improved Wilson fermions constructed from the recently suggested nHYP links\cite{Hasenfratz:2007rf}
are stable even at a coarse lattice spacing. The additional cost of the smearing
is small and more than compensated by the improved conditioning number of the fermion matrix.
Exploratory studies of non-perturbative improvement of this action using Schr{\"o}dinger functional
techniques have also been presented at this conference~\cite{Roland07}.

\section{Smearing procedure}
Most dynamical algorithms are based on molecular dynamics and therefore
one needs to differentiate the action with respect to the gauge fields.
If one uses smeared links, the derivative of the smeared link with
respect to the thin link is needed. This turns out to be a problem 
for the projection which is part of the definition of the APE smearing,
\begin{equation}
V_{n,\mu} ={\rm Proj}_{{\rm SU}(3)} \left[ (1+\alpha) U_{n,\mu} + 
\frac{\alpha}{6
}\sum _{\pm \nu \neq \mu }
U_{n,\nu ;\mu }
U_{n+\hat{\nu },\mu ;\nu }
U_{n+\hat{\mu },\nu ;\mu }^{\dagger }\right] \ ,
\end{equation}
with $n$ labeling the site and the result of the projection $B={\rm Proj}_{{\rm SU}(3)} A$ 
is defined as the matrix $B\in {\rm SU}(3)$ which maximizes ${\rm tr} [A^\dagger B+B^\dagger A]$.
The stout smearing of Morningstar and Peardon \cite{Morningstar:2003gk} 
is a fully differentiable alternative. However, we found it to be
considerably less effective in reducing the effect of the dislocations.

It also turns out that typically one level of smearing is not enough and
one needs to iterate the procedure. However, iterating it too many times
leads to Dirac operators with a large footprint.
In quenched studies, a particular smearing recipe has proved to be efficient but not
over-doing it: HYP smearing\cite{Hasenfratz:2001hp}. It consists of three
levels of projected APE smearing, however, the smearing is restricted such
that the smeared link only receives contributions from its hypercube. 
In Ref.~\cite{Hasenfratz:2007rf} we introduced n-HYP smearing which differs
from the original HYP smearing only in that the projection is not to SU$(3)$
but to U$(3)$
\begin{eqnarray}
V_{n,\mu }&=
&{\rm Proj}_{U(3)}[(1-\alpha _{1})U_{n,\mu }+
\frac{\alpha _{1}}{6}\sum _{\pm \nu \neq \mu }\tilde{V}_{n,\nu ;\mu }\tilde{V}_{n+\hat{\nu },\mu ;\nu }\tilde{V}_{n+\hat{\mu },\nu ;\mu }^{\dagger }]\,  \nonumber ,\\ 
\tilde{V}_{n,\mu ;\nu }&=&{\rm Proj}_{U(3)}[(1-\alpha_{2})U_{n,\mu }
+\frac{\alpha_{2}}{4}\sum _{\pm \rho \neq \nu ,\mu }\bar{V}_{n,\rho ;\nu \, \mu }\bar{V}_{n+\hat{\rho },\mu ;\rho \, \nu }\bar{V}_{n+\hat{\mu },\rho ;\nu \, \mu }^{\dagger }]\, ,
\\
\bar{V}_{n,\mu ;\nu \, \rho }&=&{\rm Proj}_{U(3)}[(1-\alpha _{3})U_{n,\mu }+\frac{\alpha _{3}}{2}\sum _{\pm \eta \neq \rho ,\nu ,\mu }U_{n,\eta }U_{n+\hat{\eta },\mu }U_{n+\hat{\mu },\eta }^{\dagger }] \, , \nonumber
\end{eqnarray}
where  $V_{n, \mu}$ is the link from which the Dirac operator is to be constructed.
The projection is defined by
\begin{equation}
{\rm Proj}_{U(3)} A = A \frac{1} {\sqrt{A^\dagger A}}
\end{equation}
which is differentiable where $A$ is non-singular.

The details of how the derivative is computed are described  in
Ref.~\cite{Hasenfratz:2007rf}. The technique is very similar to the 
one use for stout smearing and results in similar computational cost.
In principle, the discontinuity where the matrix $A$ is singular 
could cause problems in the simulation, however, in practice this 
turns out not to be the case in the simulations presented below.
Apart from the simulations described in these proceedings, n-HYP smearing
has also been used in overlap simulations~\cite{DeGrand:2007tm,Hasenfratz:2007iv,Anna07}.

\section{The simulation}
In order to test the performance of the smeared link, we started
simulations of two degenerate flavors of improved Wilson fermions on $16^3\times32$
lattices. We use tadpole improved L\"uscher--Weisz gauge action. The
clover coefficient is set to its tree level value $c_{SW}=1$. All
links of this operator are constructed from n-HYP links, where the
coefficients are set to the standard HYP values, $\alpha_1=0.75$, 
$\alpha_2=0.6$ and $\alpha_3=0.3$.

We generated lattices at two values of the sea quark mass. Both
runs give a Sommer parameter $r_0/a=3.8$ which, using $r_0=0.5$fm, translates
to a lattice spacing of $a\approx0.13$fm.
One ensemble, labeled heavy in the following---with $\beta=7.2$, $\kappa=0.1272$---has
a pseudo-scalar mass of  about 520MeV . A lighter run
with $\beta=7.1$, $\kappa=0.1280$ renders $m_{\rm PS}\approx 360$MeV.
We use multiple time scale Hybrid Monte Carlo\cite{Urbach:2005ji} with Hasenbusch's mass
preconditioning\cite{Hasenbusch:2001ne}. We typically ran at 90\% acceptance rate 
at moderate step size and unit length trajectories. 

We note in passing, that the overhead in the computation associated with
the nHYP smearing---construction of the smeared links and the additional
differentiation---is modest. For the light $16^3\times32$ run 13\% of 
the wall clock time is spent on this part whereas we spend two third of the
time on the inversions and 18\% on the gauge force. We believe that the 
cost of the nHYP smearing is more than off-set by the improvement of 
the condition number of the fermion matrix and the reduced cost of the inversions.

\subsection{Auto-correlation}

\begin{figure}
\begin{center}
\includegraphics[width=0.55\textwidth]{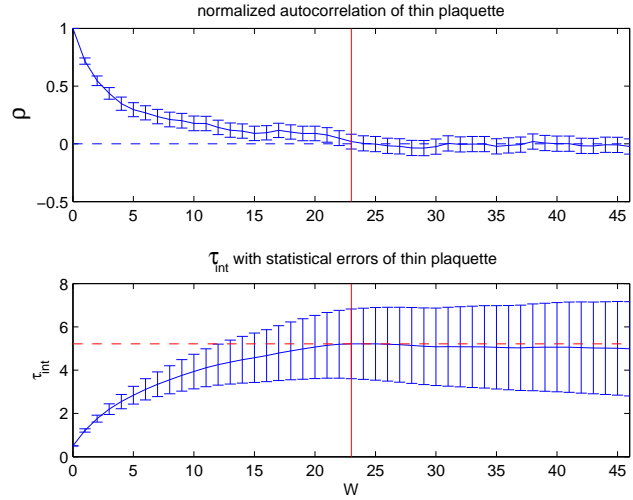}
\end{center}
\caption{\label{fig:ac}Auto-correlation analysis of the plaquette for the light
run on the $16^3\times32$, $\beta=7.1$, $\kappa=0.1280$ lattice.}
\end{figure}
In order to make sure that using smeared links does not lead to increased
auto-correlation times, we compute $\tau_{\rm int}$ for the plaquette and
the plaquette constructed from smeared links which is less UV dominated. We
use the methods of Ref.~\cite{Wolff:2003sm}.
For the light ensemble, we
find $\tau_{\rm int}=5.2(1.4)$ and $10(4)$ respectively, see Fig.~\ref{fig:ac}. 
The heavy runs with $m_{\rm PS}\approx 520$MeV have $\tau_{\rm int}=6(2)$
and $9(3)$, which are surprisingly similar (if one ignores the large errors). 
However, the parameters of the
algorithm have been tuned more carefully for the light run which might
explain the absence of critical slowing down.

\begin{figure}
\begin{center}
\includegraphics[width=0.3\textwidth,angle=-90]{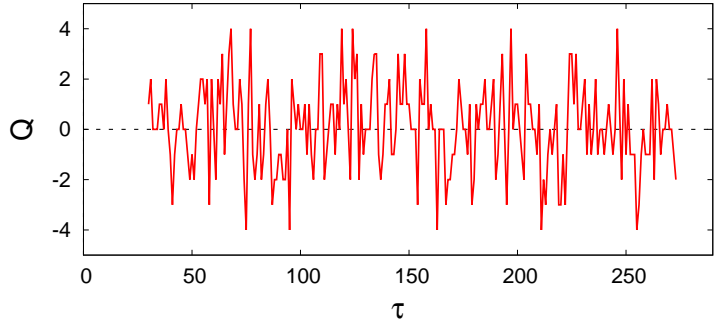}
\end{center}
\caption{\label{fig:topo}Monte-Carlo history of the topological charge. Each
unit on the x-axis corresponds to five trajectories of unit length 
by which the measurements are separated.}
\end{figure}

In particular in simulations with the DBW2 gauge action, the auto-correlation
time of the topological charge frequently are  hundreds of 
trajectories. We test for this on a smaller lattice, $12^3\times24$, $L\approx1.4$fm and
$m_{\rm PS}\approx 450$MeV. We measure the topological charge as defined by 
the index of the overlap operator after every fifth trajectory. The result
is shown in Fig.~\ref{fig:topo}. We find no detectable sign for increased auto-correlation.

We take these two findings as evidence that the intuitive expectation that the
smeared links do not have any negative effects on the auto-correlation time is 
indeed confirmed.

\subsection{Stability}

In order to get reliable results from a simulation one needs to be sure
that it is stable. How to check for stability is matter of some debate.
We have tried to look for signs of meta-stabilities as put forward by Refs.~\cite{Farchioni:2004us} and \cite{Farchioni:2004fs}
and the distribution of the lowest eigenvalue of the Hermitian Dirac
operator advocated in Refs.~\cite{Debbio:2005qa} and \cite{Debbio:2006cn}.

First to the meta-stabilities: the method is to compare observables (typically the 
plaquette) as a function of Monte--Carlo time from a stream which is 
started from a hot configuration with a stream which starts from a 
unit, cold configuration.
\begin{figure}
\begin{center}
\includegraphics[width=0.3\textwidth,angle=-90]{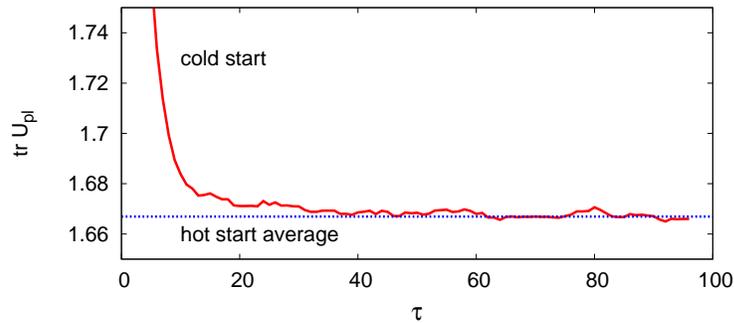}
\end{center}
\caption{\label{fig:plaq}Monte-Carlo time history of the plaquette from a 
cold start. The horizontal line denotes the average from the hot-start.
There is no sign for instabilities from a first-order phase transition}
\end{figure}
The result for the light run is shown in Fig.~\ref{fig:plaq}. We observe
that the cold start reaches the average of the plaquette from the 
hot start after about 50 trajectories and there is no sign for intermediate
meta-stabilities.

Since this is reassuring, we can now turn to the distribution of the lowest
magnitude eigenvalue of the Hermitian Dirac operator $Q=\gamma_5 D$.
Ref.~\cite{Debbio:2005qa} uses this distribution as the main
indicator for the stability of a simulation. Due to the singularity in
the force where $Q$ has a zero eigenvalue, instabilities occur once
these surfaces are crossed too often during the generation of the ensemble.
A situation where the distribution of the lowest eigenvalue is far 
away from zero (where far is defined in units of the width of the 
distribution) is interpreted as stable. 

\begin{figure}
\begin{center}
\includegraphics[width=0.3\textwidth,angle=-90]{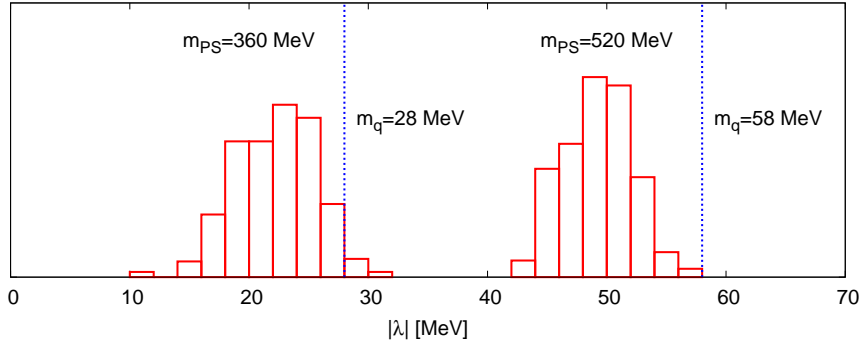}
\end{center}
\caption{\label{fig:gap}Distribution of the modulus of the lowest eigenvalue
of the Hermitian Dirac operator $\gamma_5 D$ for the two large volume runs.}
\end{figure}

In Fig.~\ref{fig:gap} we have plotted the distribution of $|\lambda_0|$, 
the magnitude of the eigenvalue of $Q$ with the  smallest absolute value.
It contains the result for the light and the heavy run along with its
bare PCAC quark mass.  For the heavy ensemble, we find a mean of the 
distribution  $\mu=49$MeV and a width of $\sigma=3$MeV, the width is defined as in 
Ref.~\cite{Debbio:2005qa}. The light run has a median gap of 23MeV and
also a width of 3MeV. As argued in Ref.~\cite{Debbio:2005qa} the width
is independent of the quark mass. Also both of our runs fulfill $3 \sigma < \mu$,
the criterion of stability given in that paper. If one assumes that $\sigma$
is independent of the sea quark mass, and also assumes that the lower bound
of stability is given by this criterion, one gets that a pion mass of 
240MeV is reachable with our setup without running into problems with
the stability.
As a side remark, we notice that the bare quark mass is quite close to the 
median of the gap, however, some unknown renormalization constants are
needed for a quantitative comparison.

\begin{figure}
\begin{center}
\includegraphics[width=0.3\textwidth,angle=-90]{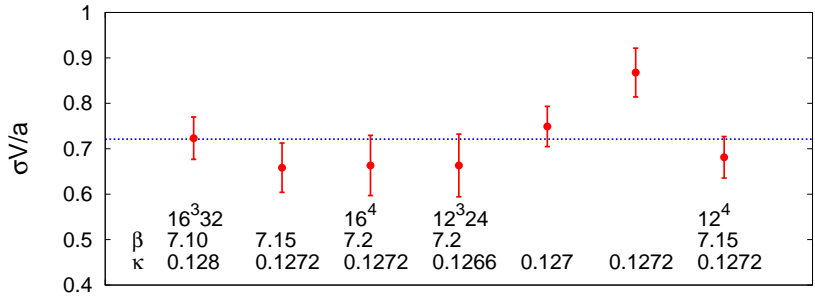}
\end{center}
\caption{\label{fig:width}Scaling of the width $\sigma$ of the distribution
of  the lowest eigenvalue of $Q$ for various volumes and lattice spacings.}
\end{figure}

In the same paper\cite{Debbio:2005qa} it has been argued that the combination
$\sigma V/a$ should be a scaling quantity. Our result for a selection
of our runs is displayed in Fig.~\ref{fig:width}. We find a value of about 0.7
whereas the original publication\cite{Debbio:2005qa} found for thin
link unimproved Wilson fermions values around 1. However, since we use the
bare width $\sigma$ in this plot, a meaningful comparison is again not possible.

\section{Summary}
At this conference, we presented new results from simulations with
nHYP smeared improved Wilson fermions in 2fm boxes at a coarse lattice 
spacing of about 0.13fm and light pion mass $m_{\rm PS}\approx 360$MeV.
We showed that the simulations are stable in this region where ordinary
thin link Wilson simulations are at least very difficult.
Contrary to special  gauge actions (e.g. DBW2) which suppress dislocations, smeared links have no
negative impact on auto-correlation times.  Given the experience
of quenched simulations we also expect an improved scaling behavior. This
however, is subject of future studies.

If one assumes the stability bound of Ref.~\cite{Debbio:2005qa} to be 
correct, we could simulate at pion masses down to 240MeV without running
into problems with the stability. This would correspond to a 
$m_{PS} L \approx 2.4$. From the scaling arguments of the same paper one
would conclude that all reasonable parameter values in the p-regime 
are accessible even at such a coarse lattice spacing.

\section{Acknowledgments}
We thank the Zeuthen computer center of DESY for providing us with access to their
linux clusters.  This research was partially supported by the US Dept. of Energy.

\end{document}